\documentclass[10pt,twocolumn]{revtex4}
\usepackage{graphicx,epsfig,amsmath,amsthm,amssymb}
\usepackage[colorlinks=true, pdfstartview=FitV, linkcolor=blue, citecolor=blue, urlcolor=blue]{hyperref}

\begin{document}

\title{Building Thinking Skills in Pre-service Elementary School Teachers}

\author{Nathan Moore}
\affiliation{Winona State University, Department of Physics}
\email{nmoore@winona.edu}
\author{Jacqueline R. O'Donnell}
\affiliation{Rochester STEM Academy, Science Department, Rochester MN}
\email{jodonnell@rochesterstemacademy.com}
\date{\today}

\begin{abstract}
The work describes curricular modification to a physical science content class for elementary education majors at 
Winona State University
[a regional comprehensive university].   
The modification, the regular use of Cognitive Acceleration materials to develop student reasoning ability, produces unusually high gains on  Lawson's Classroom Test of Scientific Reasoning.  
\end{abstract}

\maketitle

\section{Investigative Science at Winona State}
Elementary Education majors at 
Winona State University  
[a regional comprehensive university]
take two 4-credit courses in  science to fulfill their general education laboratory science requirements.  The courses, Investigative Science 1 \& 2, cover licensing standards in conceptual Physics, Chemistry, Earth Science, and Biology.  The courses are formulated to present science content in the way we hope our children are taught at the elementary level: always in the lab, always experiencing data before learning formal concepts, and always coming to consensus about the meaning of data through group discussion and metacognition.  The approach is meant to reduce trepidation toward science by providing experience with inexpensive activities which can be easily done in the students' future classrooms. 

The courses are taught by a mix of faculty from across the college of science and engineering. While not a new approach to teaching introductory science, \cite{1970_cohort,fullerton,Rice_GeosEd,science_101,inquiry_connections}, the classes, run regularly for the past 7 years, have been sufficiently successful to be the required science classes for elementary education students on campus.

Historically, our progress in teaching this course has been measured with internally developed content exams and also by Anton Lawson's ``Classroom Test of Scientific Reasoning,'' (CTSR) \cite{Lawson_CTSR}, a multiple choice instrument which suggests a student's Piagetian reasoning level.  The CTSR has been shown to be a meaningful predictor of success on learning measures in physics, such as the Force Concept Inventory \cite{CTSR_and_FCI}.  The measure quantifies old suspicions about our students, \cite{renner_teaching}, and in general, we consider gain on the CTSR an indicator that the course is changing, for the better, the way our pre-service teachers think about the world.  

\section{The Classroom Test of Scientific Reasoning}
The CTSR is a 24 item test that is graded out of 13 points, as most of the questions are paired in the style of ``Given this situation, what will happen?" and ``Why?".  Students must get both parts of each pair correct to score a point, and the measure is accordingly quite difficult.  Scores of 0-4 correspond to concrete reasoners, 5-10 transitional reasoners, and 11-13 formal.  The transitional period is often broken into early (5-7) and late (8-10) transitional reasoning groups.  For an overview of these reasoning levels, see \cite{Lawson_2000}.  

As children advance through primary school, the topics and skills discussed move from concrete operations (eg ``What does the letter `A' look like?", ``How many words are in this sentence?") towards formal logical operations (eg ``Are all squares also rectangles?", ``Where in a sentence does the verb go?", ``Why are many of a bird's bones hollow?", ``If a chicken has 4 fingers, is their number 10 the same as our number 10?").  If a teacher lacks the ability to reason formally, it is likely that they have always learned science by memorization.  It is unreasonable then to expect such a teacher to ever fully address the ``why" part of a scientific explanation adequately without communicating the model building and testing ability that building an effective scientific explanation requires.    

According to the psychologist Jean Piaget, \cite{piaget_1966},
students should begin to transition to formal reasoning (i.e., the students don't require a physical object in front of them (concrete reasoning), but instead are capable of performing thought experiments (formal reasoning)) around the age of sexual maturity, age 11 in girls and 12 in boys.   Our local measurements with the CTSR suggest that either Piaget was optimistic or standards have slipped, see Table \ref{tab:ctsr} for example scores from one of the author's (Moore's) 
classes.
These data corroborate with many other studies that have found relatively small percentages of students that have acquired formal reasoning, \cite{lawson_1985}

Historically speaking, according to this measure only a modest fraction of students in the education program enter as formal reasoners.  This reasoning ability is necessary if students are going to sum the forces on the international space station, imagine the motion of  $N_2$ molecules within a balloon, or hypothesize about the effect of a drug on a population of bacteria.  Our fields quickly reduce to a set of disconnected rules and special cases for the students so lacking.  Further, the ability to create a controlled experiment (hypothetical-deductive reasoning, for example ``if-and-then-therefore" statements) is very difficult for students not at the  ``formal" reasoning level, \cite{lawson_2002}.
It should be apparent that future teachers must be equipped with these skills.

As mentioned, table \ref{tab:ctsr} lists CTSR pre-scores and gains for a number of classes at Winona State University.
At present, the authors are aware of no published normal distributions of score or gain for reformed/traditional science classes on this measure.  Locally, it seems that a ``typical" gain on the measure is about 1 point, and apocryphally, many science classes show gains of zero.  This may seem odd, but the CTSR ``measures" scientific reasoning ability, which develops either in response to stimulation or as a part of normal maturation, which recently has been observed to be delayed \cite{delayed_maturation}. 
  
  \section{Cognitive Acceleration Interventions}
At the middle school level, Philp Adey, Michael Shayer, and Carolyn Yates have developed supplementary lessons which stimulate the development of reasoning ability, \cite{thinking_science}.  Their results in the field of ``Cognitive Acceleration" over the past twenty years are, without hyperbole, tremendous, \cite{adey_references}, and have culminated in a series of curricular interventions appropriate for students in US grades K-6.  The original series of Cognitive Acceleration interventions, ``Thinking Science," \cite{thinking_science}, are a series of 30 ``intervention lessons" which are designed to be used with 11-14 year old children.  The lessons are intended to be inserted into the regular curriculum at the rate of once every week or two over a period of about one to two years.  They are specifically designed to create the cognitive conflict which stimulates the development of students' reasoning ability.  
The specific skills targeted by the materials are: control of variables, classification, ratio and proportionality, inverse proportionality and equilibrium, probability and correlation, and the use of abstract models to explain and predict.
For further discussion of the material's effect in the classroom, see \cite{pink_book,adey_references}.

\begin{table}[h]
\begin{center}
\begin{tabular}{ c c c c }
\hline
\textbf{Investigative Science} & &  average & average \\
Semester  & N &  pre CTSR &  CTSR gain\\
\hline
Spring 2008   & 26 & 5.5 &  1.3 \\ 
Spring 2009  & 26 & 5.9 & 0.8 \\
\textbf{Spring 2011}  & \textbf{28} & \textbf{6.4} & \textbf{2.2} \\
University Average  & 160 & 6.25 & 1.13 \\
\hline
\textbf{College Physics} &&& \\
Summer 2009 & 32 &8.7 & 1.0 \\
Summer 2011  & 26 & 8.4 & 1.5 \\
Fall 2009 & 11 & 7.5 & 1.8 \\
Fall 2010 & 19 & 7.5 & 1.2 \\
\hline
\end{tabular}
\caption{\label{tab:ctsr}Typical results from Lawson's ``Classroom Test of Scientific Reasoning", administered at 
Winona State University.
Except for the WSU average listed, all results posted are those of the author (Moore). 
The  WSU average gains in Investigative Science do not include the Spring 2011 class, the difference being the motivation for this work.
The $N$ given is the number of students who were present to take both pre and post tests, and class averages are taken only over students with a complete pre/post pair of scores.
The Spring 2011 section made use of half of the ``Thinking Science" lessons, which are designed to stimulate the development of reasoning ability among middle school students.  As can be seen from the data, they are also appropriate for use among college students. 
}
\end{center}
\end{table}

\section{Experiment and Results}
In the Spring 2011 semester, one of the authors, Moore, 
taught a section  of Investigative Science 1, which discusses ideas from Physics and Chemistry.  Having recently become aware of the Cognitive Acceleration materials, he taught the class with the normal curriculum, with one small modification.  Each weekend, students worked through one ``Thinking Science" lesson, \cite{thinking_science,pink_book}, with (substantially metacognitive) discussion on the following monday.  Students were directed to work through the lesson as students.  Further they were also to identify the main learning goal or outcome from the lesson and create  additional activities aligned with the lesson, to be used if the pupils they were working with finished early.  Students were also asked for written reflections on the nature of the learning in each lesson.  The long-term goal for the effort was for this class to be able to facilitate Thinking Science lessons with a group of 5th grade students, which sadly because of logistics, didn't materialize.  

Over the course of the semester, the pre-service elementary teachers were affected by the lessons.  Terms from the materials, like ``fair test" from the control of variables lesson, regularly began to appear in normal class discussions outside of the time allotted to Thinking Science.  In addition, in the time devoted to Thinking Science materials, student resistance to work on the lessons lessened as the semester progressed.  Group and class discussions became more in depth throughout the semester (at the beginning they would provide one word answers and try to get it over with).
This was a remarkable result, as historically this specific population has been the least willing to embrace the identity and practice of scientists.

The most substantial result from the intervention is the gain in CTSR score of about 2.2 over the course of the semester.  This is roughly double the normal increase seen in students of this ability and college major.  If this increase can be sustained in further sections of Investigative Science, it will be a powerful addition to the curriculum. 

\section{Conclusion}   
The gains in CTSR score described in this work, which came as a direct result of student exposure to Thinking Science interventions, should be seen as compelling evidence that CTSR score (and implicitly, reasoning ability) can be affected by classroom practice.  When training future educators, reasoning ability must be targeted for improvement, as teachers are not able to communicate  material beyond their own level of intellectual sophistication.  The Thinking Science materials are one way to increase this critically important ability. It would be interesting to know if the Thinking Science interventions described in the work have a similar impact on other populations with immature reasoning ability, for example, underprepared students in University (calculus-based)Physics.

\section{Acknowledgements}
The authors thank John Deming and Philip Adey for useful discussions.  The effort was supported in part by Winona State University's Teach21 initiative.

\end{document}